\documentclass[aps,pre,floats,twocolumn,showpacs,superscriptaddress]{revtex4}

\usepackage{natbib}
\usepackage[utf8]{inputenc}

\usepackage{amsmath, amsthm, amssymb}
\usepackage{enumitem}
\usepackage{graphics,graphicx}
\usepackage{epsfig}
\usepackage{times}
\usepackage{dcolumn}
\usepackage{url}
\usepackage{color}
\usepackage{array}
\usepackage{pifont}
\usepackage[all]{xy}
\usepackage{mdwlist}
\usepackage[normalem]{ulem}

\begin{document}

\newcommand{\pin}{p_\mathrm{in}}
\newcommand{\pout}{p_\mathrm{out}}

\renewcommand*\thesection{\arabic{section}}
\newcommand{\beq}{\begin{equation}}
\newcommand{\eeq}{\end{equation}}
\newcommand{\sss}{\scriptscriptstyle}
\newcommand{\sz}{\scriptsize}
\newcommand{\rev}[1]{{\color{blue} #1}}
\newcommand{\rerev}[1]{{\color{red} #1}}
\newcommand{\bea}{\begin{eqnarray}}
\newcommand{\eea}{\end{eqnarray}}
\newcommand{\nn}{\nonumber \\}
\definecolor{MyColorito}{rgb}{1,0.4,0.8}
\definecolor{MyColorito2}{rgb}{0.1,0.4,1}
\definecolor{MyColorito3}{rgb}{0.1,0.1,0.1}
\definecolor{red}{rgb}{1,0.1,0.1}
\newcommand{\ClaraMetaComment}[1]{\textcolor{MyColorito}{[\textbf{CLARA:} #1]}}
\newcommand{\clara}[1]{\textcolor{MyColorito2}{#1}}
\newcommand{\changes}[1]{\textcolor{black}{#1}}
\newcommand{\tachao}[1]{\sout{\changes{#1}}}
\definecolor{RowColor}{rgb}{0.65,0.8,0.89}
\definecolor{topColor}{rgb}{0.21, 0.10, 0.60}
\definecolor{midColor}{rgb}{0.44, 0.17, 0.53}
\definecolor{botColor}{rgb}{0.53, 0.23, 0.17}
\newcolumntype{a}{>{\columncolor{RowColor}}c}
\newcommand{\topC}[1]{\textcolor{topColor}{#1}}
\newcommand{\midC}[1]{\textcolor{midColor}{#1}}
\newcommand{\botC}[1]{\textcolor{botColor}{#1}}

\newcommand{\comment}[1]{{}}
\newcolumntype{R}[1]{>{\raggedright\let\newline\\\arraybackslash\hspace{0pt}}m{#1}}
\newcolumntype{C}[1]{>{\centering\let\newline\\\arraybackslash\hspace{0pt}}m{#1}}

\newcommand{\ba}[1]{\begin{array}{#1}}
\newcommand{\ea}{\end{array}}
\newcommand{\ben}{\begin{enumerate}}
\newcommand{\een}{\end{enumerate}}
\newcommand{\bit}{\begin{itemize}}
\newcommand{\eit}{\end{itemize}}
\newcommand{\bde}{\begin{description}}
\newcommand{\ede}{\end{description}}

\newcommand{\ds}{\displaystyle}
\newcommand{\ts}{\textstyle}
\newcommand{\st}{\scriptstyle}
\newcommand{\lra}{\longrightarrow}
\newcommand{\lmt}{\longmapsto}
\newcommand{\sign}{\mbox{sign}}
\newcommand{\req}[1]{(\ref{#1})}
\newcommand{\average}[1]{{\left\langle {#1} \right\rangle}}
\newcommand{\pderiv}[2]{\frac{\partial {#1}}{\partial {#2}}}
\newcommand{\jump}[1]{\raisebox{1ex}{\sz {#1}}}
\newcommand{\jumpu}[1]{\raisebox{1ex}{\underline{\sz {#1}}}}
\newcommand{\BbbR}{{\mathbb R}}
\newcommand{\BbbN}{{\mathbb N}}
\newcommand{\avg}[1]{\langle {#1} \rangle}
\newcommand{\NMI}{\mbox{{\em NMI}}}
\newcommand{\NVI}{\mbox{{\em NVI}}}

\title{Benchmark model to assess community structure in evolving networks}

\author{Clara Granell}
\affiliation{Departament d'Enginyeria Inform\`atica i Matem\`atiques, Universitat Rovira i Virgili, 43007 Tarragona, Spain}
\author{Richard K. Darst}
\affiliation{Complex Systems Unit, Department of Computer Science, Aalto University School of Science, P.O. Box 12200, 00076 Aalto, Finland}
\author{Alex Arenas}
\affiliation{Departament d'Enginyeria Inform\`atica i Matem\`atiques, Universitat Rovira i Virgili, 43007 Tarragona, Spain}
\affiliation{Institut Catal\`a de Paleoecologia Humana i Evoluci\'o Social, 43007 Tarragona, Spain}
\author{Santo Fortunato}
\affiliation{Complex Systems Unit, Department of Computer Science, Aalto University School of Science, P.O. Box 12200, 00076 Aalto, Finland}
\author{Sergio G\'omez}
\affiliation{Departament d'Enginyeria Inform\`atica i Matem\`atiques, Universitat Rovira i Virgili, 43007 Tarragona, Spain}

\begin{abstract}
  Detecting the time evolution of the community structure of networks is crucial to identify major changes in the internal organization of many complex systems, which may undergo important endogenous or exogenous events. This analysis can be done in two ways: considering each snapshot as an independent community detection problem or taking into account the whole evolution of the network. In the first case, one can apply static methods on the temporal snapshots, which correspond to configurations of the system in short time windows, and match afterwards the communities across layers. Alternatively, one can develop dedicated dynamic  procedures, so that multiple snapshots are simultaneously taken into account while detecting communities, which allows us to keep memory of the flow. To check how well a method of any kind could capture the evolution of communities, suitable benchmarks are needed.  Here we propose a model for generating simple dynamic benchmark graphs, based on stochastic block models. In them, the time evolution consists of a periodic oscillation of the system's structure between configurations with built-in community structure. We also propose the extension of quality comparison indices to the dynamic scenario.
\end{abstract}

\pacs{
89.75.Fb, 
89.75.Hc  
}

\maketitle

\section{Introduction}

The analysis and modeling of temporal networks has received a great deal of attention lately, mainly due to the increasing availability of time-stamped network datasets~\cite{kovanen11,holme12,perra12,starnini12,barrat13}. A relevant issue is whether and how the community structure of networks~\cite{fortunato10} changes in time. Communities reveal how networks are organized and function, hence major changes in their configuration might signal important turns in the evolution of the system as a whole, possibly anticipating dramatic developments such as rapid growth or disruption.

Indeed, there has been a great deal of activity around this topic in recent years~\cite{hopcroft04,chakrabarti06,
palla07,
ferlez08,
nussinov09,
mucha10,granell11,nussinov12,
bassett13,brodka13,ros15}. However, most investigations lack strength on the validation part, which typically consists in checking whether the results of the algorithm ``make sense'' in one or more real networks whose community structure is usually unknown. Actually, it is not obvious what exactly it means to test an algorithm for detecting evolving communities. One idea could be that of correctly identifying the community structure of the system at each time stamp. However, during the evolution of the system several events that affect the network structure may occur, such as the  creation or deletion of nodes or links or link rewiring, and it is not possible to detect these events by observing a single time-stamped network, they require taking into account the whole picture to be properly understood.

To explicitly keep track of the history of the system, an option is to consider multiple snapshots at once. For instance, in the evolutionary clustering approach~\cite{chakrabarti06} the goal is to find a partition that is descriptive of the structure of a given snapshot as well as correlated to the structure of the previous snapshots. Furthermore, the added value of any approach should be the ability to promptly detect changes in the community structure of the network. It would be possible to verify this if there were suitable benchmark graphs with evolving clusters, but those are still missing. This paper aims at filling this gap. We propose a model, derived from the classic stochastic block models~\cite{holland83,girvan02,lancichinetti08,guimera09}, that generates three classes of dynamic benchmark graphs. The objective is to provide time-evolving networks, such that at each snapshot the partition into communities is well defined according to the model. To keep things simple we consider a periodic evolution such that the same history repeats itself in cycles and is invariant under time reversal. The analysis of the community structure evolution for the designed benchmarks reveals that approaches exploiting the flow of system configurations might be more accurate in detecting the evolving community structure than methods that consider the snapshots independently. Note that in real data sets this evolution can be sharp and bursty, however in these cases the challenge of finding the community structure is not well defined,
because the range of timescales makes the mesoscopic structure clearly disconnected.

The paper is structured as follows. In Section 2 we describe the model to generate the benchmark networks, Section 3 introduces measures of comparison between dynamic clusterings, Section 4 shows an example of the application of a dynamic multislice algorithm on the proposed benchmarks. Section 5 gives a summary and reports our conclusions.

\section{Model description}

The model we propose for generating networks with evolving community structure is based on
the classic stochastic block model (SBM)~\cite{holland83}. It works as
follows. A network is divided into a number $q$ of subgraphs and the
nodes of the same subgraph are linked with a probability $\pin$,
whereas nodes of different subgraphs are linked with a probability
$\pout$. Such probabilities match the link densities within and
between subgraphs. Supposing subgraphs of equal size, if
$\pin > (q-1)\pout$ the resulting subgraphs are
communities, as the (expected) link density within subgraphs exceeds
their connectivity to the rest of the graph.
The generation of samples from
this model has a built-in efficiency: If there are $m_\mathrm{max}$
pairs of nodes, the actual number of edges is drawn from a binomial
distribution with parameters $m_\mathrm{max}$ and $p$.  Then, we
simply place this number of edges randomly to generate a sample from
our ensemble.

The model implements the two fundamental classes of dynamic processes:
growing or shrinking and merging or splitting of communities. By combining
these two reversible types of processes one can capture the most
common behaviors of dynamic communities in real systems.  We are then able to generate
three standardized benchmarks: One consists in communities that grow
and shrink in size (keeping fixed the total number of nodes of the
network), while the second considers communities that merge and
split. The third one is a mixed version of the previous two, and
consists of a combination of the last four operations.

\subsection{Grow-shrink benchmark}

This process models the movement of
nodes from one community to another.  At all times, two communities
are kept in a SBM ensemble with intracommunity link density
$p_\mathrm{in}$ and intercommunity link density $p_\mathrm{out}$.
However, the number of nodes in the two communities changes over time.
In the basic process, we have a total of $2n$ nodes in two
communities.  In the balanced state, these are split into two equal
communities of $n$ nodes, which we call $A$ and $B$.  At the extremes, a fraction $f$ of nodes in
community $A$ will switch to community $B$.  If we take
$n_\mathrm{l}$ as the size of community $A$, then the
number of nodes in the community $B$ is $n_\mathrm{r} =
2n-n_\mathrm{l}$.  Then, at time $t$ the number of nodes in
community $A$ is
\begin{equation}
  \label{eq:nleft}
  n_A = n - n f \left[  2x(t+\tau/4)-1  \right]
\end{equation}
with the $\tau/4$ phase factor specifying equal sized
communities at $t=0$.
The function $x(t)$ is the triangular waveform
\begin{equation}
  \label{eq:x}
  x(t) = \left\{
      \begin{array}{l@{\hspace{2em}}r@{\hspace{1ex}}c@{\hspace{1ex}}l}
       2t^*,      & 0   & \leq  t^*  & < 1/2 \\
       2 - 2t^*,  & 1/2 & \leq  t^*  & < 1
      \end{array}
    \right.
\end{equation}
(with $t^* \equiv (t/\tau + \phi) \bmod 1$),
which controls the time periodicity.  The constant $\phi$ is a phase
factor with $\phi=0$ for the $q=2$ case and specified otherwise in
the case of $q>2$.  With this formulation, we get communities of
sizes $(n, n)$, $(n-nf, n+nf)$, $(n, n)$, and $(n+nf, n-nf)$ at
$t/\tau \bmod 1=0$, ${1 \over 4}$, ${2 \over 4}$, and ${3 \over 4}$,
respectively.  In practice, all $2n$ nodes are sorted in some
arbitrary order, and the first $n_A$ nodes are put into
community $A$, and the others into
community $B$.  Say these nodes are $i=0$ to $i=2n-1$.

After the community sizes are decided, the edges must be placed, taking into account that it
is necessary that we keep the two communities in the proper SBM ensemble with equal and
independent link probability $p_\mathrm{in}$ at all times.  The independence of pairs
provides a hint on how to do this.  When a node $j$ is moved from community $A$ to $B$, all the
existing edges of node $j$ are removed. Then an edge is added between $j$ and each node in
the destination community $B$ with equal and independent probability $p_\mathrm{in}$ and between
$j$ and each node in community $A$ with equal and independent probability $p_\mathrm{out}$,
thus the ensemble is maintained.  Conveniently, all edges can be pre-computed and stored to
allow a strictly repeating process, with the state at time $t$ being identical to the state at
time $t+\tau$, in analogy to the merging process.

\begin{figure}
  \centering
   \includegraphics*[width=.45\textwidth]{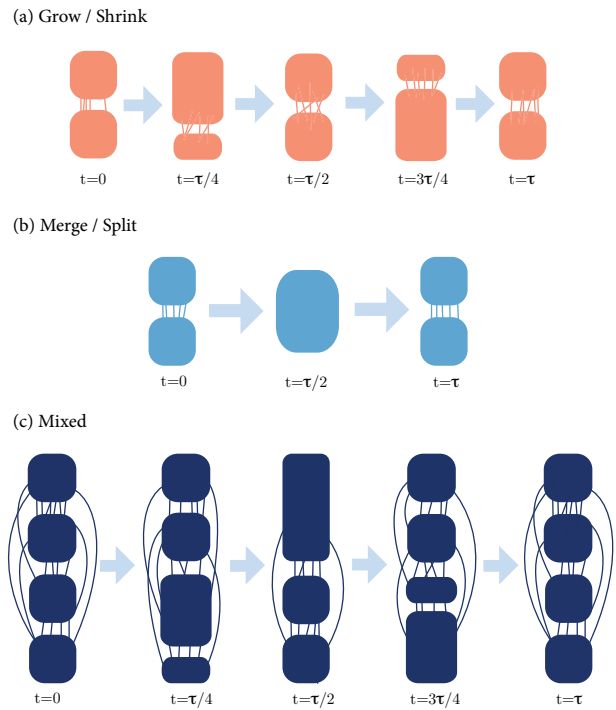}
   \caption{(Color online) Schematic representation of the benchmarks.  (a)
     Grow-shrink benchmark with $q=2$.  We begin with two equal-sized
     communities, and over a period of $\tau$ nodes move from the
     bottom community to the top, then from the top to bottom, then
     back to the symmetric state.  (b) Merge-Split benchmark with
     $q=2$.  We begin with two communities, and over a period of
     $\tau$ we linearly add edges until there is one community with
     uniform link density, then reverse the process. (c) Mixed
     benchmark with $q=4$, combining the merging and growing
     processes.}
  \label{fig:modelfig-q2}
\end{figure}

A special case that we need to cope with is the situation where $f$ is very high and $\pin$ is
very low. When this happens, a community shrinks too
much and it may become disconnected. In order to preserve the ensemble,
we do not take actions to totally eliminate this possibility, but we ensure that $n(1-f)\pin
\gg 2$ to reduce the probability of disconnection. However, if a disconnection
occurs, the process is aborted and re-run. Figure~\ref{fig:modelfig-q2}(a) is a sketch of the grow-shrink benchmark for the case $q=2$.

\subsection{Merge-split benchmark}

This process models the merging of two communities.  In this setup, we have a set of
$2n$ nodes, divided into two communities of $n$ nodes each.  Each of the
two initial communities has a link density of $p_\mathrm{in}$, where
those links are placed at initialization and kept unmodified over time. There are two extreme states: the unmerged and the merged state.
In the unmerged state, all possible pairs of nodes between the two communities have an edge with probability
$p_\mathrm{out}$. This means that the network still has a connected component, but the nodes form two communities. In the merged state, all possible pairs of nodes between these two communities have an edge with probability
$p_\mathrm{in}$, which implies that all pairs of nodes in the network have the same link
density $p_\mathrm{in}$, the previous two communities are now indistinguishable, and thus we have one large community with
$2n$ nodes.

The merge-split process is a periodic interpolation of the merged and
unmerged states.  The numbers of intercommunity edges in the unmerged
state $m_\mathrm{um}$ and in the merged state $m_\mathrm{m}$ are
first picked from a binomial distribution consistent with the binomial
distribution parameters $n^2$ and $p_\mathrm{out}$ or $p_\mathrm{in}$.
All possible intercommunity edges are placed in some arbitrary but
random order, and the first
\begin{equation}
  \label{eq:mstar}
  m^*(t) = [1-x(t)]m_\mathrm{um} + x(t)m_\mathrm{m}
\end{equation}
edges are selected to be active at time $t$.  The effective
intra-community link density is $p_\mathrm{inter}^*(t) = m^*(t)/n^2$.
The parameter $x(t)$ is the triangular waveform from Eq.~(\ref{eq:x}).
In practice, this means that at time
$t/\tau \bmod 1 = 0$ the communities are unmerged and at $t/\tau
\bmod 1 = 1/2$ the communities are merged, with linear interpolation
(of the number of edges) between these points.  Since the possible
edges are ordered only at initialization, the process is strictly
periodic, that is, the edges present at time $t$ are identical to
those present at time $t+\tau$.

One may think that the communities are fully merged at the extreme of
this process, where the intercommunity link density is
$p_\mathrm{inter}^* = p_\mathrm{in}$ (at $t=\tau/2$).  However, due to
the \textit{detectability limit} of communities in stochastic
block models, this is not the case~\cite{decelle11}. Even when $p_\mathrm{out}
< p_\mathrm{in}$, it can be that the configuration is
indistinguishable from one large community.  Following \cite{decelle11}, at the
point
\begin{equation}
  p_\mathrm{in} - p_\mathrm{inter}^*
  = \sqrt{ \frac{1}{n} \left( p_\mathrm{in}+p_\mathrm{inter}^* \right) }
\end{equation}
we consider the communities to be merged into one for all practical
purposes.  While this limit is strictly speaking only accurate in the
sparse and infinite-size limit, it is an adequate approximation.
A schematic representation of the merge-split benchmark, for $q=2$ is shown in Fig.~\ref{fig:modelfig-q2}(b).

\subsection{Mixed benchmark}

This process is a combination of the merging and growing processes. In this process, there is a total of
$4n$ nodes with two merging-splitting communities ($2n$ nodes) and two
growing-shrinking communities ($2n$ nodes). The intra-community
links are managed with the same processes as above with phase factors
of $\phi=0$ for both. If there are $q=4a>4$ total communities, then
the pairs of communities involved in merging and growing process have phase
factors $\phi=0, {1\over a}, {2\over a}, ... {{a-1} \over a}$.
Between the pairs of nodes that belong to different processes, an edge exists with a probability of $p_\mathrm{out}$.
Figure~\ref{fig:modelfig-q2}(c) exemplifies the mixed benchmark when
$q=4$.


\section{Time-dependent comparison measures}

The assessment of the performance of any clustering algorithm requires
the use of measures to define the distance or similarity between any
pair of partitions. The list of available measures is long, including
e.g.\ the Jaccard index \cite{jaccard1912}, the Rand index
\cite{rand71}, the adjusted Rand index \cite{hubert85}, the normalized
mutual information \cite{strehl02}, the van Dongen metric
\cite{vandongen00} and the normalized variation of information metric
\cite{meila07}. All of them have in common the possibility of being
expressed in terms of the elements of the so-called confusion
matrix or contingency table, thus we focus first on its
calculation. Let ${\cal C}=\{C_{\alpha}|\alpha=1,\ldots,r\}$ and
${\cal C'}=\{C'_{\alpha'}|\alpha'=1,\ldots,r'\}$ be two partitions of
the data in $r$ and $r'$ disjoint clusters. The $\alpha\alpha'$th component of the
contingency table $M$ accounts for the number of elements in the
intersection of clusters $C_{\alpha}$ and $C'_{\alpha'}$,
\begin{equation}
  m_{\alpha\alpha'}=|C_{\alpha}\cap C'_{\alpha'}|\,.
  \label{maap}
\end{equation}
The sizes of the clusters simply read
$n_{\alpha}=|C_{\alpha}|=\sum_{\alpha'} m_{\alpha\alpha'}$ and
$n'_{\alpha'}=|C'_{\alpha'}|=\sum_{\alpha} m_{\alpha\alpha'}$ and the
total number of elements is $N=\sum_{\alpha} n_{\alpha}=\sum_{\alpha'}
n'_{\alpha'}=\sum_{\alpha}\sum_{\alpha'}m_{\alpha\alpha'}$.
With these
definitions at hand, one can calculate the Jaccard index,
\begin{equation}
  J = \frac{\ds \sum_{\alpha}\sum_{\alpha'} \binom{m_{\alpha\alpha'}}{2}}
           {\ds \sum_{\alpha} \binom{n_{\alpha}}{2}
              + \sum_{\alpha'} \binom{n'_{\alpha'}}{2}
              - \sum_{\alpha}\sum_{\alpha'} \binom{m_{\alpha\alpha'}}{2}
           }\,,
  \label{jaccard}
\end{equation}
the normalized mutual information index,
\begin{equation}
  \NMI = \frac{\ds -2\sum_{\alpha}\sum_{\alpha'} m_{\alpha\alpha'} \log \frac{N m_{\alpha\alpha'}}{n_{\alpha} n'_{\alpha'}}}
              {\ds \sum_{\alpha} n_{\alpha} \log \frac{n_{\alpha}}{N}
                 + \sum_{\alpha'} n'_{\alpha'} \log \frac{n'_{\alpha'}}{N}
              }\,,
  \label{nmi}
\end{equation}
and the normalized variation of information metric,
\begin{equation}
  \NVI = \frac{-1}{\log N} \sum_{\alpha}\sum_{\alpha'} \frac{m_{\alpha\alpha'}}{N}
           \log \frac{(m_{\alpha\alpha'})^2}{n_{\alpha} n'_{\alpha'}}\,,
  \label{nvi}
\end{equation}
where, by convention, $0\log 0 = 0$.

\begin{figure}[tb]
  \centering
   \includegraphics*[width=.375\textwidth]{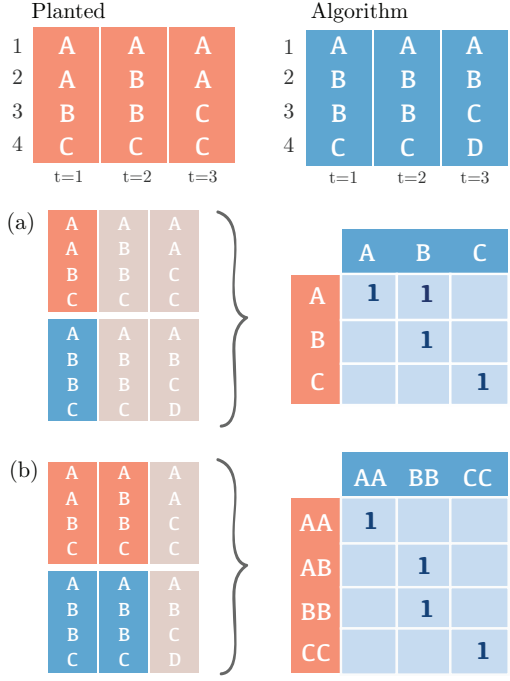}
   \caption{(Color online) Construction of the contingency tables $m_{\alpha\alpha'}$.
   On top we represent three steps (columns) of the time evolution of a network of four nodes (rows),
   and the partitions in communities we want to compare, e.g.\ the planted partitions from the
   benchmark and those obtained by a certain algorithm. To compare these two partitionings,
   we can do it as it is depicted in (a), which takes only one snapshot at a time ($\sigma=0$),
   or as in (b), building a contingency table where the entries consider two snapshots at the
   same time ($\sigma=1$). Afterward, the measures (NVI, NMI or Jaccard index) are calculated from these tables.}
  \label{fig:measures}
\end{figure}

In the case of evolving networks we have to compare two sequences of
partitions $\{{\cal C}(t)|t=1,\ldots,T\}$ and $\{{\cal
  C'}(t)|t=1,\ldots,T\}$, a task that can be performed in different
ways. The simplest solution is the independent comparison of
partitions at each time step, by measuring the similarity or distance
between ${\cal C}(t)$ and ${\cal C'}(t)$ for each value of $t$, thus
obtaining ,e.g., a Jaccard index $J(t)$ for each snapshot, see Fig.~\ref{fig:measures}(a).
However,
this procedure discards the evolutionary nature of the communities: We
would like to quantify not only the static resemblance of the
communities but also if they evolve in a similar way.

Our proposal consists in the definition of windowed forms of the different
indices and metrics, obtained by considering sequences of consecutive
partitions, i.e.\ time windows of a predefined duration $\sigma$. In Fig.~\ref{fig:measures}(b) we show
the comparison between individual snapshots and sequences of length 2. For example, let us consider the time window formed by time steps from $t$
to $t + \sigma$. Every node belongs to a different cluster at each
snapshot, and this evolution can be identified as one of the items in
${\cal D}(t;\sigma)={\cal C}(t)\times{\cal
  C}(t+1)\times\cdots\times{\cal C}(t+\sigma)$ for the first sequence of
partitions, and ${\cal D'}(t;\sigma)={\cal C'}(t)\times\cdots\times{\cal
  C'}(t+\sigma)$ for the second one, where the multiplication sign denotes the Cartesian product of sets. Since the number of nodes is
$N$, there are at most $N$ different nonvoid sets
$D_{\alpha}(t;\sigma)\in {\cal D}(t;\sigma)$ and the same for
$D'_{\alpha'}(t;\sigma)\in {\cal D'}(t;\sigma)$.
For example, in Fig.~\ref{fig:measures}(b), the combinations of partitions (excluding empty sets) are ${\cal D}(t=1;\sigma=1)=\{ \mbox{\sc aa}, \mbox{\sc ab}, \mbox{\sc bb}, \mbox{\sc cc}\}$ and ${\cal D'}(t=1;\sigma=1)=\{ \mbox{\sc aa}, \mbox{\sc bb}, \mbox{\sc cc}\}$.
Next, we may define
the elements of the contingency table for this time window as
\begin{equation}
  m_{\alpha\alpha'}(t;\sigma)=|D_{\alpha}(t;\sigma)\cap D'_{\alpha'}(t;\sigma)|\,,
  \label{maapw}
\end{equation}
which accounts for the number of nodes following the same
cluster evolutions $D_{\alpha}(t;\sigma)$ and
$D'_{\alpha'}(t;\sigma)$. Likewise, we have
\begin{align}
  n_{\alpha}(t;\sigma)&=|D_{\alpha}(t;\sigma)|=\sum_{\alpha'} m_{\alpha\alpha'}(t;\sigma)\,,
  \\
  n'_{\alpha'}(t;\sigma)&=|D'_{\alpha'}(t;\sigma)|=\sum_{\alpha} m_{\alpha\alpha'}(t;\sigma)\,,
\end{align}
and
\begin{equation}
  N=\sum_{\alpha} n_{\alpha}(t;\sigma)=\sum_{\alpha'} n'_{\alpha'}(t;\sigma)=\sum_{\alpha}\sum_{\alpha'}m_{\alpha\alpha'}(t;\sigma)\,.
\end{equation}
Finally, we may use Eqs.~(\ref{jaccard})--(\ref{nvi}) to calculate
the corresponding windowed Jaccard index $J(t;\sigma)$, windowed
normalized mutual information index $\NMI(t;\sigma)$, and windowed
normalized variation of information metric $\NVI(t;\sigma)$,
respectively. Of course, the windowed measures reduce to the standard
static ones when $\sigma=0$, and are able to capture differences
in the evolution of communities that cannot be distinguished using their
classical versions (see the Appendix).

We will see in the next section how the plots of $\NVI(t;\sigma)$ are valuable to
compare different algorithms and to detect in which moments of the time evolution
they differ. Nevertheless, it is also convenient to have a single number to quantify the
overall deviation. A simple solution is the use of the average squared errors, which
is expressed as follows:
\begin{eqnarray}
  E_{\mbox{\sz J}}(\sigma) & = & \frac{1}{T}\sum_{t=1}^T [J(t;\sigma)-1]^2\,,
  \\
  E_{\mbox{\sz \NMI}}(\sigma) & = & \frac{1}{T}\sum_{t=1}^T [\NMI(t;\sigma)-1]^2\,,
  \\
  E_{\mbox{\sz \NVI}}(\sigma) & = & \frac{1}{T}\sum_{t=1}^T \NVI(t;\sigma)^2\,.
  \label{nvierr}
\end{eqnarray}

For simplicity and for its superior mathematical properties (see \cite{meila07})
we have chosen to use only the NVI metrics in the rest of this article.
See Supplemental Material for the results using the normalized mutual information and the Jaccard index.


\section{Results}

Here we show an example of the application of a community detection algorithm,
designed to take into account the evolution of complex networks, to reveal the community
structure in our benchmarks. The chosen method is the
multislice algorithm in \cite{mucha10}, which extends the
definition of modularity to multilayer networks.
In their representation, each layer (slice)
consists of a single network at a particular time. The slices are
connected between them by joining each node with its counterpart in
the next and previous layer, and this link has a specified weight~$\omega$,
equal for all links of this kind, which acts as a tuning parameter.
For $\omega=0$, no connection between slices is considered and the algorithm is
performed statically. As this value increases, more consideration is
given to the communities across layers. The formulation includes an
additional parameter~$\gamma$, which accounts for the tuning of the
resolution at which communities are found, in the manner of~\cite{reichardt04}. In
this work, we have used the code available in~\cite{genlouvain_site},
setting the resolution parameter~$\gamma$ to 1 and varying the
interslice coupling~$\omega$.

\begin{figure*}[ht]
 \begin{center}
    \mbox{\includegraphics*[height=.90\textwidth,angle=-90]{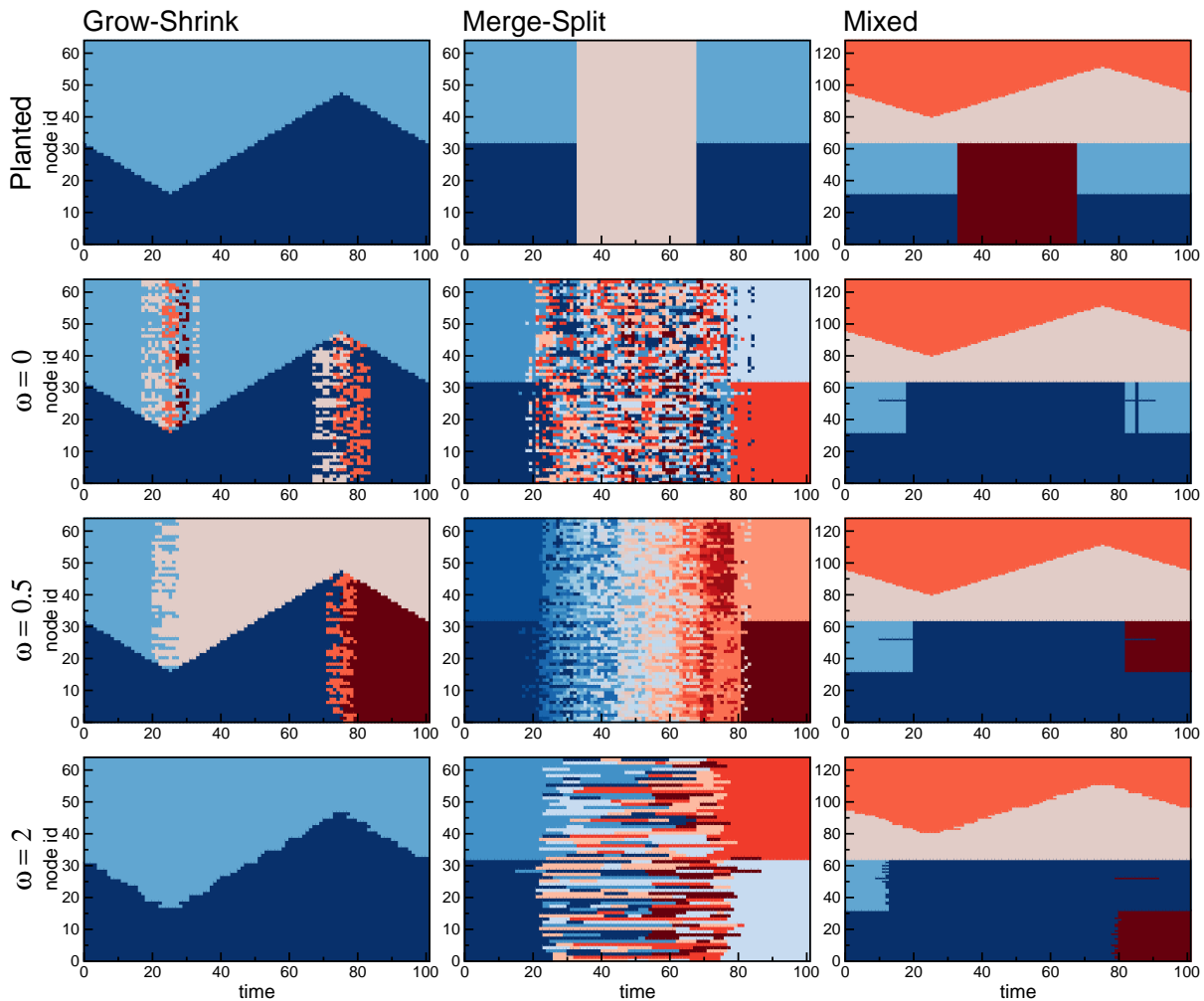}}
  \end{center}
  \caption{(Color online) Results of the application of the multislice community detection
  method to the three benchmarks proposed (in columns). The first row corresponds to the planted
  partition of each benchmark, while the three remaining rows are the partitions obtained by the
  multislice algorithm for different values of the interslice parameter $\omega$, which is the weight
  of the coupling between different instances of the same nodes across layers.
  When $\omega=0$ the slices are
  disconnected and then the community detection analysis is done for each slice separately. As this
  value increases, more importance is given to the evolving nature of the problem, and communities
  across slices are found. In each plot, the vertical axis corresponds to the index of nodes in the
  network, while the horizontal axis represents the time. The color of each pair \{node, time\} is
  the label of the community at which the node is assigned at that specific time.
    }
  \label{fig:partitions}
\end{figure*}

The benchmarks used to put to test this algorithm are generated
using the model proposed in this paper. For the sake of simplicity,
we generate three simple standard benchmarks, one for each basic
procedure: grow-shrink, merge-split and
mixed. The {grow-shrink benchmark consists in a
network with $q=2$ communities, where each community has initially
$n=32$ nodes (therefore the total size of the network is $N=64$), with
$\pin=0.5$, $\pout=0.05$, $f=0.5$, and $\tau=100$ time steps. The
merge-split test has a variable number of communities; in
this paper we use the parameters $q=2$ communities of size $n=32$
each, with $\pin=0.5$,  $\pout=0.05$, and $\tau=100$. The mixed benchmark,
a combination of the previous two, has $q=4$ communities of $n=32$ nodes each,
and the other parameters are set as in the previous cases.

Figure~\ref{fig:partitions} shows the planted partitions for the
three benchmarks and the results from the multislice algorithm at three
different interslice couplings:
In the extreme case $\omega=0$ slices are
considered independently, $\omega=0.5$ is an intermediate value that provides good results, and
$\omega=2$ provides an example of the partitioning obtained when using
strong coupling between layers. \changes{It can be seen that for $\omega=0$ we obtain a different partition for each time step, and the results are mostly correct, except for those configurations of the sizes of the communities where the preference of modularity for equal-sized communities hampers the process (see the first column of Fig.~\ref{fig:partitions}). Higher values of $\omega$ request higher consistency through time, which implies that the number of misclassified individual snapshots is reduced.}
We have also compared the multislice method with a
temporal stability approach \cite{petri2014} and the results obtained
are very similar to the results of the multislice algorithm obtained at $\omega=0.5$.


\begin{figure*}[ht]
 \begin{center}
    \mbox{\includegraphics*[width=.90\textwidth]{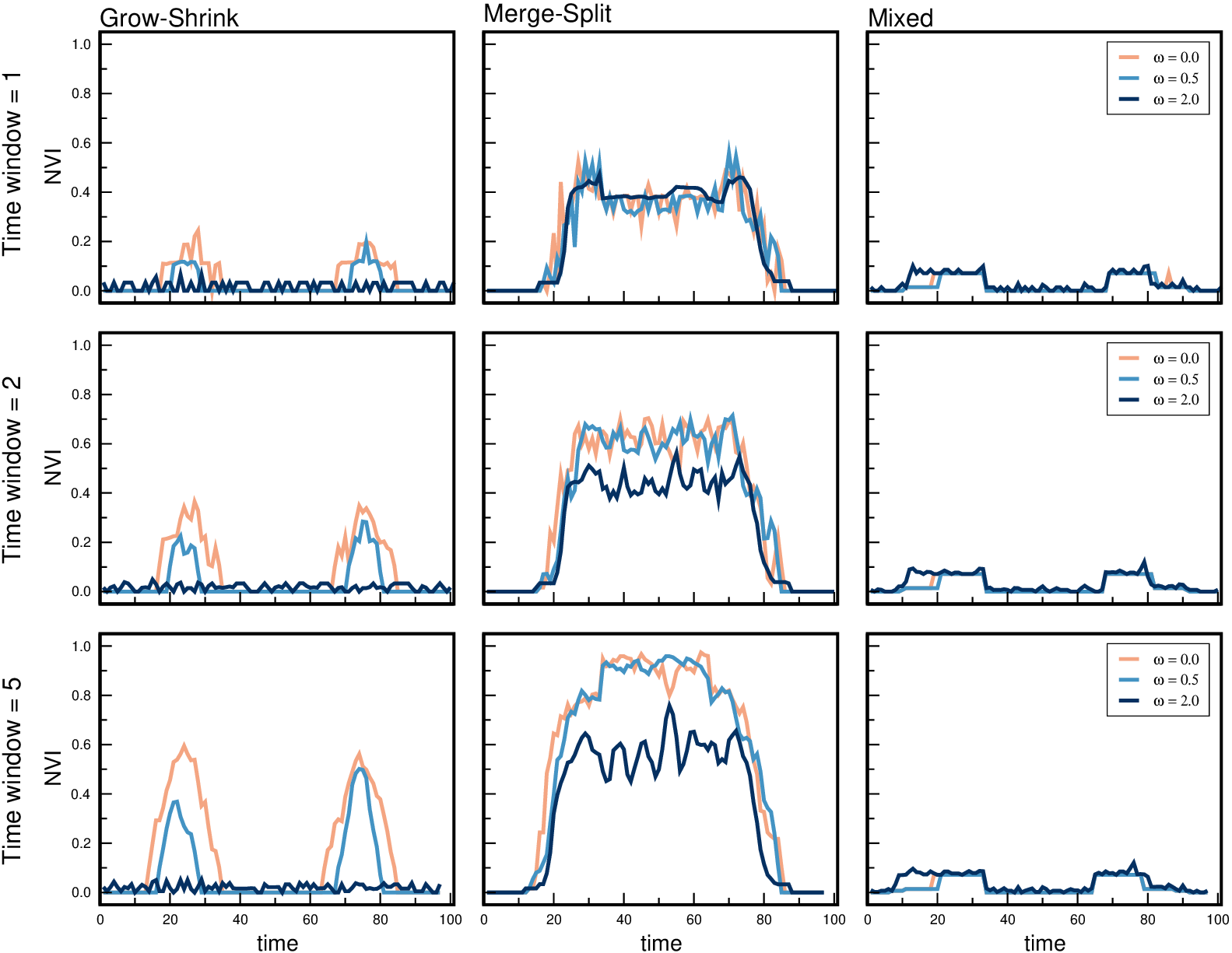}}
  \end{center}
  \caption{(Color online) Plots of the normalized variation of information (NVI) between
    the planted partition and the results of the multislice algorithm in Fig.~\ref{fig:partitions},
    for three different interslice couplings and for the three benchmarks proposed. The NVI is computed
    using the proposed evolving formulation and for three different window sizes: 1, 2 and 5. There
    is a column for each benchmark, and a row for each time window size.}.
  \label{fig:nvi}
\end{figure*}

To quantitatively evaluate the results, we
use the windowed measures introduced in the previous section. We calculate the
measures between the partitions obtained by the algorithm and the
planted ones, for three values of the time window. When the time window
is of size 1 ($\sigma=0$), each snapshot is considered independently,
that is, we have computed the measure between the planted partition at $t$ and
the algorithm's result at $t$, repeating this process until $t=\tau$.
Instead, with the time window of size 2 ($\sigma=1$), we evaluate the evolution of the partitions
during two consecutive time steps, following the same process but comparing the planted
partitions at $[t,t+1]$ with the algorithm's results at $[t,t+1]$. This formulation is more
restrictive, as we impose, in addition to the condition that the
nodes must belong to the same community, that their evolution during two consecutive time steps is
also the same. Similarly, we have also analyzed time windows of size 5 ($\sigma=4$) to check
the quality of the detected community evolutions at longer ranges.

Figure~\ref{fig:nvi} shows the results for the NVI.
We observe that, for the grow-shrink benchmark, the error is large for
$\omega=0$, but becomes almost zero at $\omega=2$. Moreover, the values of the NVI increase
with the size of the time window for $\omega=0$ and $\omega=0.5$, but in a larger amount
when the parameter corresponds to the static version of the multislice algorithm. This
means that the interslice weight is helping to find the persistence of nodes in their communities,
as expected. The merge-split benchmark shows an almost identical bad performance for
the three values of~$\omega$ at windows of size~1, but $\omega=2$ does not make it worse when the
size of the window increases, unlike the other two. The mixed benchmark
is quite neutral, with just a small difference from $\omega=2$. Finally, the NVI squared
errors reported in Table~\ref{tab:nvierr} and calculated using Eq.~(\ref{nvierr}) are in
perfect agreement with this analysis. The results using the NMI and Jaccard indices (see
the Supplemental material) also support these observations. Thus, we may conclude that,
in this case, the use of memory to track the evolution of communities is convenient, but
the trade-off between the continuity of the community structure and its static relevance
must be carefully adjusted.

\begin{table*}[t]
  \begin{center}
  \begin{tabular}{R{15mm}C{12mm}C{18mm}C{18mm}C{8pt}c}
    \hline \hline
                   & Time   &  \multicolumn{4}{c}{NVI squared error}   \\  \cline{3-6}
     Multislice    & window &  Grow-shrink &  Merge-split & & Mixed    \\  \hline \hline
                   & 1      &  0.0065      &  0.0851      & & 0.0015   \\
     $\omega=0.0$  & 2      &  0.0201      &  0.2146      & & 0.0015   \\
                   & 5      &  0.0658      &  0.4427      & & 0.0016   \\  \hline
                   & 1      &  0.0023      &  0.0808      & & 0.0014   \\
     $\omega=0.5$  & 2      &  0.0067      &  0.2019      & & 0.0014   \\
                   & 5      &  0.0242      &  0.4278      & & 0.0015   \\  \hline
                   & 1      &  0.0006      &  0.0878      & & 0.0023   \\
     $\omega=2.0$  & 2      &  0.0005      &  0.1113      & & 0.0024   \\
                   & 5      &  0.0006      &  0.1922      & & 0.0029   \\
    \hline \hline
  \end{tabular}
  \end{center}
  \caption{The NVI squared error, for each method tested and each benchmark in Fig.~\ref{fig:partitions},
  considering three different time windows.}
  \label{tab:nvierr}
\end{table*}




\section{Conclusions}

We have presented a simple model based on the stochastic block model that allows for the
construction of time-dependent networks with evolving community
structure. It is useful for benchmarking purposes in testing the
ability of community detection algorithms to track properly the
structural evolution. We have also introduced extended
time-dependent measures for the comparison of different partitions in
the dynamic case, which allow for the observation of differences between
the outcome of the algorithms and the planted partitions through time.

Our code for benchmark generation and the time-dependent comparison
indices is available at \cite{rkd_multiplex_code} and released under
the GNU General Public License.

\begin{acknowledgments}
  This work was partially supported by MINECO through Grant
  No.\ FIS2012-38266; and by the EC FET-Proactive Project PLEXMATH (Grant No.\
  317614). A.A.\ also acknowledges partial financial support from the
  ICREA Academia and the James S.\ McDonnell Foundation.  R.K.D.\ and
  S.F.\ gratefully acknowledge MULTIPLEX, Grant No.\ 317532 of the
  European Commission, and the computational resources provided by
  Aalto University Science-IT project.
\end{acknowledgments}

\section*{Appendix}

\section*{Distinguishing community evolutions with windowed measures}

Figure~\ref{fig:appendix} shows an example in which, according to the planted
partitions, the eight nodes of a network are divided in two communities of four nodes each and
these partitions remain constant throughout the three times steps of the network evolution.
Two different community detection algorithms find the communities evolutions represented in
Figs.~\ref{fig:appendix}(a) and~\ref{fig:appendix}(b), which are characterized by the assignment of just one node to the
wrong community at each time step. In Fig.~\ref{fig:appendix}(a) this node is the fourth one during the three time
steps, while in Fig.~\ref{fig:appendix}(b) they are the second, the third, and the sixth, respectively. Since the
nature of the mistake is the same at all time steps, the comparison of the planted and
algorithm partitions with a time window of size~1 generates equivalent contingency tables,
thus the standard comparison measures do not change in time, with a constant value of the
NVI equal to~0.2856. However, if we take into account a time window of size~3, the two
evolving community structures detected by the algorithms are different, yielding structurally
different contingency tables and values of the NVI equal to~0.2856 and~0.3852, respectively.
Therefore, the conclusion is that windowed measures give complementary information for the
comparison of time evolving community structures due to their capacity to take into
account several snapshots at the same time.

\begin{figure*}[ht]
  \begin{center}
    \begin{tabular}{lll}
      (a) & \hspace{1cm} (b) \\
      \mbox{\includegraphics*[width=.35\textwidth]{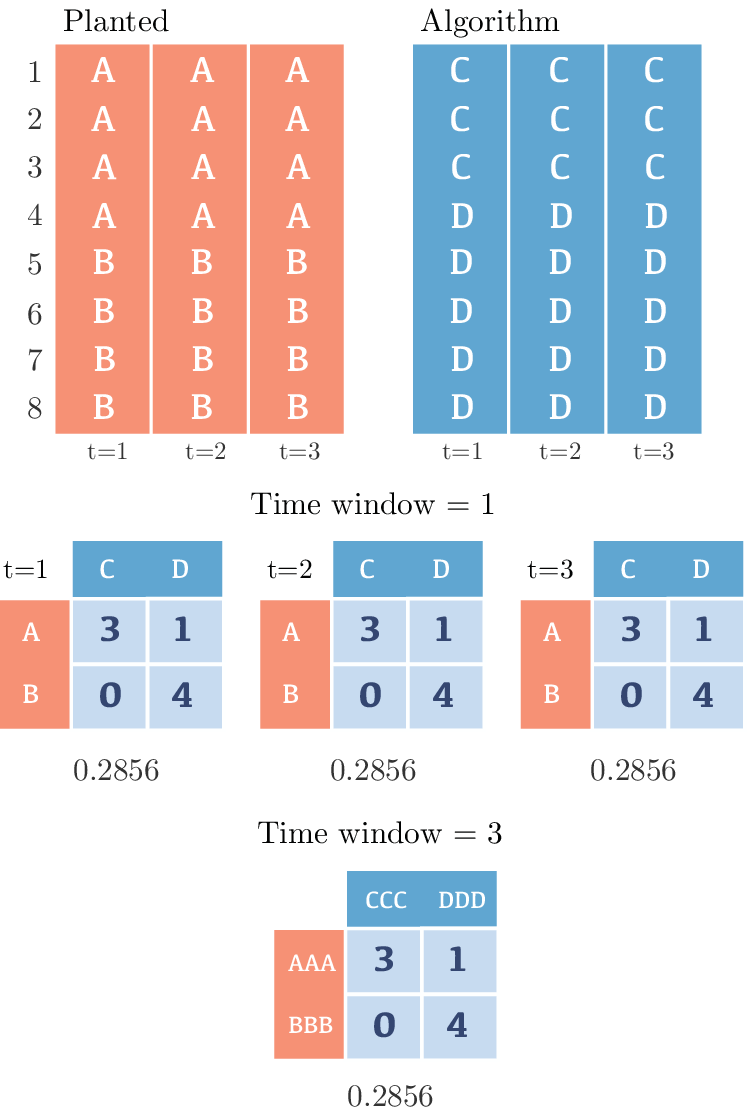}} & &
      \mbox{\includegraphics*[width=.35\textwidth]{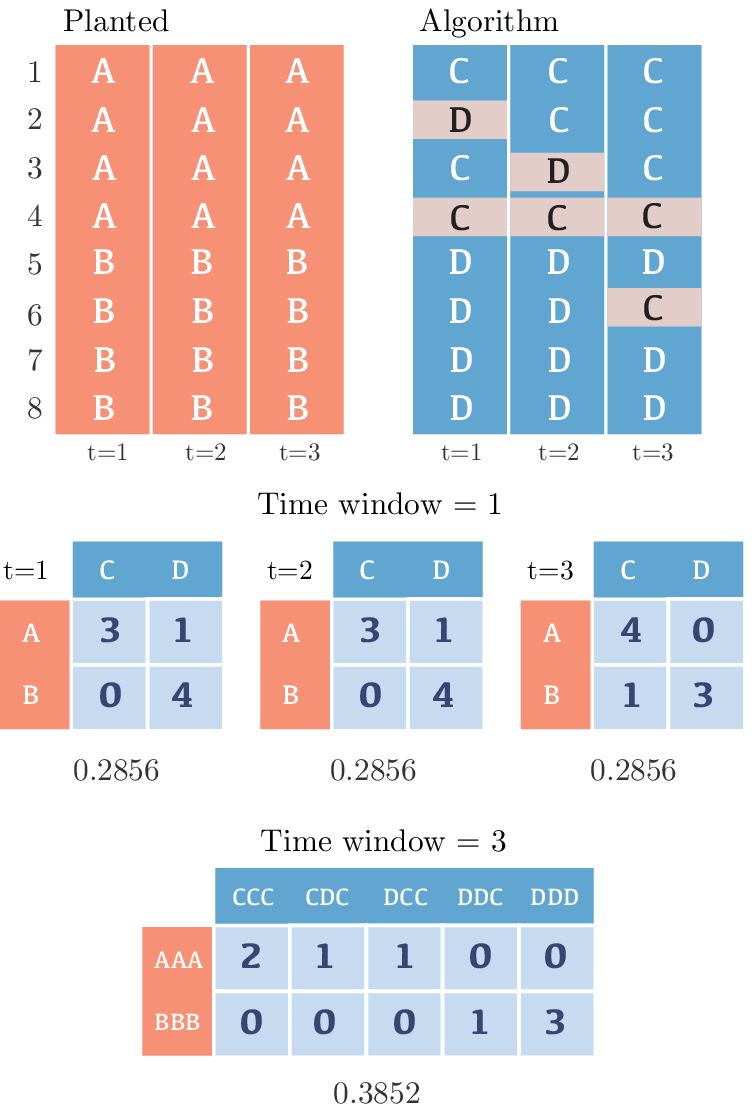}}
    \end{tabular}
  \end{center}
  \caption{(Color online) Example of the comparison of a planted evolving community
  structure with the results from two different algorithms. The NVI values are shown below the
  corresponding contingency tables for time windows of sizes~1 and~3.}.
  \label{fig:appendix}
\end{figure*}


\section*{Supplemental material}
Additional Tables~\ref{tab:jerr} and~\ref{tab:nmierr} and Figs.~\ref{fig:jacc} and~\ref{fig:nmi}.

\begin{table*}[t]
  \mbox{}
  \vspace{1cm}
  \begin{center}
  \begin{tabular}{R{15mm}C{16mm}C{23mm}C{23mm}C{8pt}c}
    \hline \hline
                  & Time   &  \multicolumn{4}{c}{Jaccard squared error} \\  \cline{3-6}
    Multislice    & window &  Grow-shrink &  Merge-split & & Mixed      \\  \hline \hline
                  & 1      &  0.0720      &  0.3345      & & 0.0307     \\
    $\omega=0.0$  & 2      &  0.1365      &  0.4840      & & 0.0303     \\
                  & 5      &  0.2336      &  0.6272      & & 0.0325     \\  \hline
                  & 1      &  0.0293      &  0.3193      & & 0.0276     \\
    $\omega=0.5$  & 2      &  0.0546      &  0.4608      & & 0.0282     \\
                  & 5      &  0.1105      &  0.6013      & & 0.0303     \\  \hline
                  & 1      &  0.0019      &  0.3326      & & 0.0360     \\
    $\omega=2.0$  & 2      &  0.0014      &  0.3605      & & 0.0374     \\
                  & 5      &  0.0147      &  0.4488      & & 0.0421     \\
    \hline \hline
  \end{tabular}
  \end{center}
  \caption{Jaccard squared error, for each method tested and each benchmark,
  considering three different time windows.}
  \label{tab:jerr}
\end{table*}

\begin{table*}[t]
  \mbox{}
  \vspace{1cm}
  \begin{center}
  \begin{tabular}{R{15mm}C{16mm}C{23mm}C{23mm}C{8pt}c}
    \hline \hline
                  & Time   &  \multicolumn{4}{c}{NMI squared error}   \\  \cline{3-6}
    Multislice    & window &  Grow-shrink &  Merge-split & & Mixed    \\  \hline \hline
                  & 1      &  0.0337      &  0.4932      & & 0.0067   \\
    $\omega=0.0$  & 2      &  0.0621      &  0.4806      & & 0.0063   \\
                  & 5      &  0.1022      &  0.4855      & & 0.0059   \\  \hline
                  & 1      &  0.0143      &  0.4896      & & 0.0060   \\
    $\omega=0.5$  & 2      &  0.0262      &  0.4753      & & 0.0059   \\
                  & 5      &  0.0479      &  0.4790      & & 0.0055   \\  \hline
                  & 1      &  0.0065      &  0.4951      & & 0.0094   \\
    $\omega=2.0$  & 2      &  0.0041      &  0.4891      & & 0.0094   \\
                  & 5      &  0.0041      &  0.4825      & & 0.0100   \\
    \hline \hline
  \end{tabular}
  \end{center}
  \caption{NMI squared error, for each method tested and each benchmark,
  considering three different time windows.}
  \label{tab:nmierr}
\end{table*}

\begin{figure*}[t]
\begin{center}
\includegraphics[width=14cm]{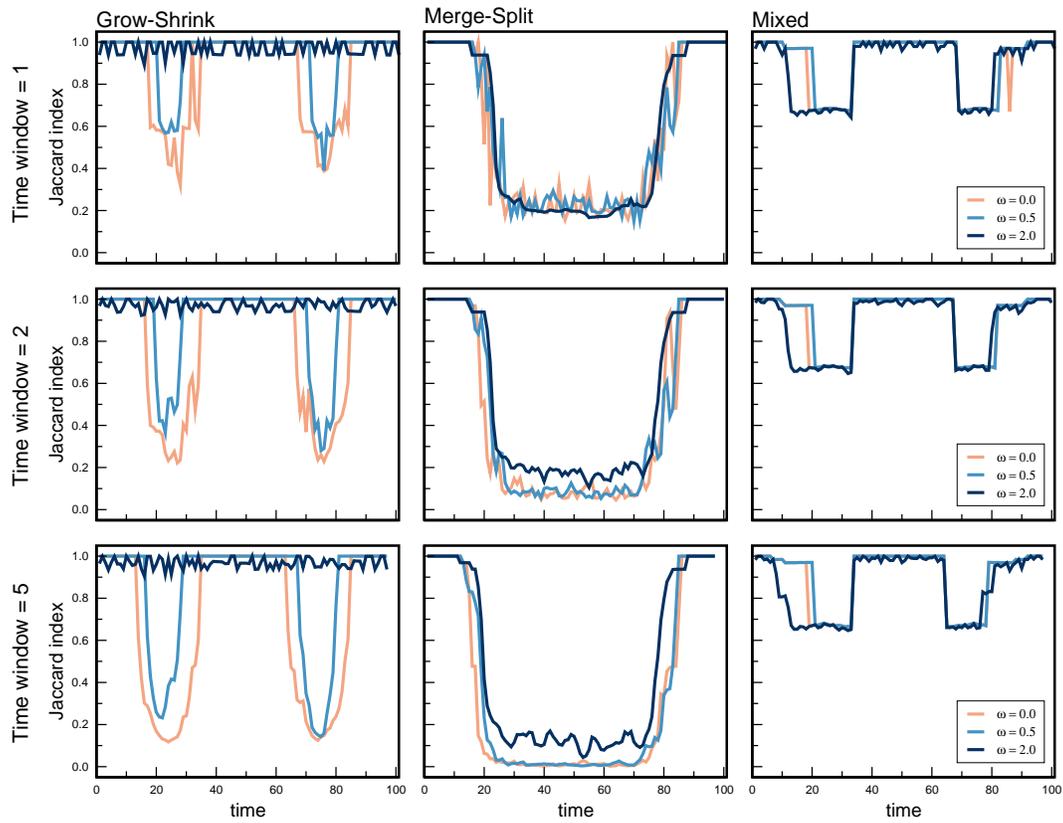}
\end{center}
\caption{Plots of the Jaccard Index between
    the planted partition and the results of the multislice algorithm for three different interslice couplings and for the three benchmarks proposed. The Jaccard index is computed
    using the proposed evolving formulation and for three different window sizes: 1, 2 and 5. There
    is a column for each benchmark, and a row for each time window size.}
    \label{fig:jacc}
\end{figure*}

\begin{figure*}[t]
\begin{center}
\includegraphics[width=14cm]{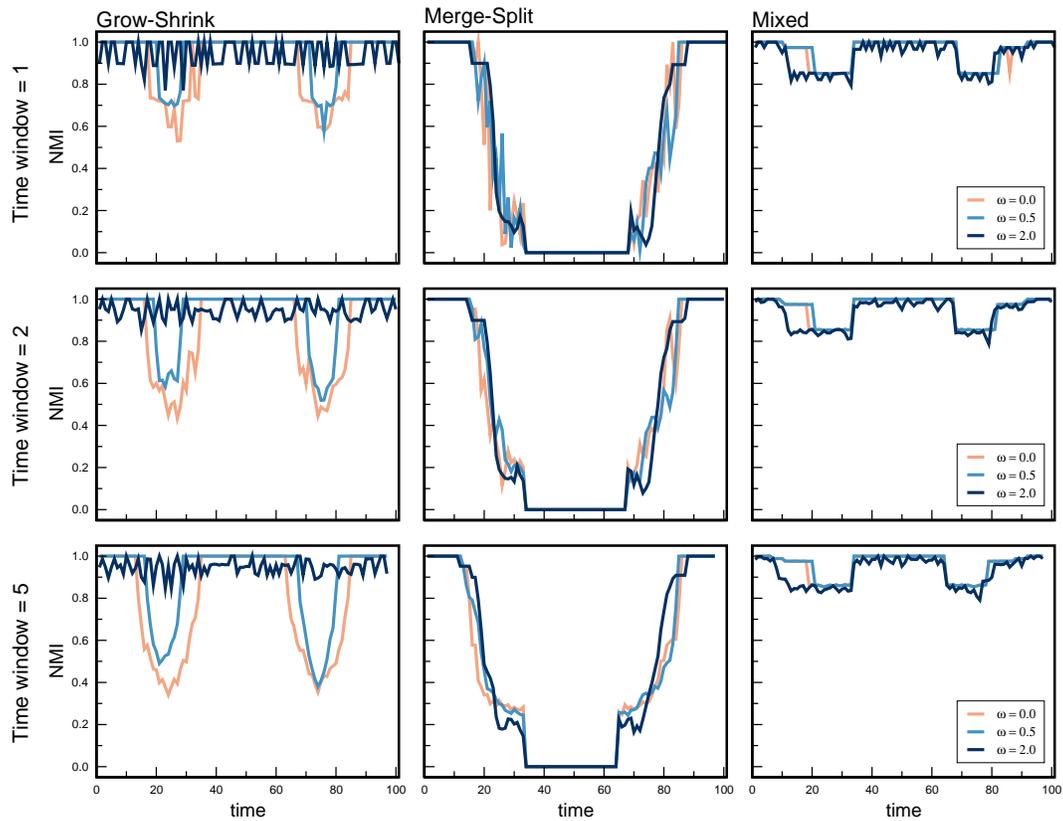}
\end{center}
\caption{Plots of the Normalized Mutual Information (NMI) between the planted partition and the results of the multislice algorithm for three different interslice couplings and for the three benchmarks proposed. The NMI is computed using the proposed evolving formulation and for three different window sizes: 1, 2 and 5. There
    is a column for each benchmark, and a row for each time window size.}
    \label{fig:nmi}
\end{figure*}

\end{document}